# Expressing and Executing Informed Consent Permissions Using SWRL: The All of Us Use Case

**Muhammad Amith, MS, PhD[1±], Marcelline R. Harris, PhD, RN, FACMI[2±], Cooper Stansbury, MS[2], Kathleen Ford, MM[2], Frank J. Manion, PhD, FAMIA[3], Cui Tao, PhD, FACMI[1*]**

[1]School of Biomedical Informatics, University of Texas Health Science Center, Houston, TX; [2]University of Michigan, Ann Arbor MI; [3]Melax Technologies, Houston TX

[±]contributed equally to this work
[*]corresponding author:cui.tao@uth.tmc.edu

**Abstract**

*The informed consent process is a complicated procedure involving permissions as well a variety of entities and actions. In this paper, we discuss the use of Semantic Web Rule Language (SWRL) to further extend the Informed Consent Ontology (ICO) to allow for semantic machine-based reasoning to manage and generate important permission-based information that can later be viewed by stakeholders. We present four use cases of permissions from the All of Us informed consent document and translate these permissions into SWRL expressions to extend and operationalize ICO. Our efforts show how SWRL is able to infer some of the implicit information based on the defined rules, and demonstrate the utility of ICO through the use of SWRL extensions. Future work will include developing formal and generalized rules and expressing permissions from the entire document, as well as working towards integrating ICO into software systems to enhance the semantic representation of informed consent for biomedical research.*

**Introduction**

Informed consent (IC) is a *process* intended to ensure that an individual (or their legally authorized representative), has sufficient information and understanding to voluntarily make a decision about participating in a research study. IC *documents* provide individuals with the information needed to make a decision about whether to volunteer for a research study and serve as a record of the decisions made during IC. In this way, IC documents serve as an important communication vehicle between the research team and potential study participants (or legally authorized representatives). IC documents provide information about actions the signer of the IC document prescribes as allowable, given the information within the IC document.

The specific research context we address are studies that aim to collect and share biospecimens, the data derived from analysis of those specimens, and other sources of data that can be associated with the study participant. Informed consent documents, when signed, establish and preserve important linkages among the persons signing the consent form, the research team, stewards and managers of specimen and data repositories, and other potential future users of specimens and/or data; the IC documents become a 'source of truth' regarding the allowability of potential actions by the research team and others.

There are several notable efforts that focus on open, computable representations of permissions expressed in IC consent documents. For example, the Community Based Collaborative Care Resource Work Group within Health Level Seven (HL7) is developing a consent resource specification called the Fast Healthcare Interoperability Resources (FHIR) at https://www.hl7.org/fhir/consent.html. The resource targets four use cases: (1) privacy consent directive, (2) medical treatment consent directive, (3) research consent directive, and (4) advance care directives. Currently, the published version (v4.01) is limited to the privacy consent directive. Of particular relevance to our interest, developers of the FHIR resource define a consent directive as "the legal record of a patient's (e.g. a healthcare consumer) agreement with a party responsible for enforcing the patient's choices, which permits or denies identified actors or roles to perform actions affecting the patient within a given context for specific purposes and periods of time." A consent form is defined as "Human readable consent content describing one or more actions impacting the grantor for which the grantee would be authorized or prohibited from performing. It includes the terms, rules, and conditions pertaining to the authorization or restrictions, such as effective time, applicability or scope, purposes of use, obligations and prohibitions to which the grantee must comply." The FHIR consent resource does not yet address the challenge of expressing the multiple consent directives that may be present in a single IC document, nor the challenge of linking detailed information such as actors, actions, and conditions to one or more consent directives within the document. Our teams have reviewed thousands of research and clinical IC documents, using both manual and machine learning based approaches[1,2]. Many if not most of the documents we reviewed include multiple permission-directives (i.e., consent directives), are often stated ambiguously, and contain relations to clauses throughout the document that are necessary to understand the permission itself. Since the FHIR specification is not fully developed, value sets that reflect the full spectrum of terminology needs related to research permissions are not yet identified.

Another example of efforts to support computable representations of permissions in IC documents is published by the Global Alliance for Genomics and Health (GA4GH). Included in their regulatory and ethics toolkit are suggested consent clauses for genomic research and for rare disease research, as well as machine readable consent guidance. The consent clauses are intended to be used as a resource for researchers as they draft IC documents, with language that is consistent with the GA4GH standards for essential consent elements. Currently, the consent clauses are available only as text-based sentences; there is no machine-processable format. The GA4GH machine readable consent guidance similarly provides another resource for researchers as they construct IC documents, however this resource provides a mapping from a set of sample consent clauses to terms [3] that have been adopted by the Data Use Ontology (DUO). DUO is published within the OBO Foundry suite of ontologies, and based on the Basic Formal Ontology that is the upper level ontology on which all OBO Foundry ontologies are built, http://www.obofoundry.org/ontology/duo.html. The intent of DUO is to standardize downstream data use concepts into a machine-readable format while tagging datasets with restrictions about their usage; DUO lacks detail about the IC process and documents.

While the HL7 and GA4GH efforts are intended to support consistent and comparable exchange of information about consent directives and potentially more detailed information about permitted actors, actions, and conditions, neither supports full semantic interoperability including IC processes and entities expressed in IC documents. A desirable feature of full semantic interoperability is that it allows for the use of semantic reasoners, based on predicate logic, to infer relationships between information; in this case, information about permissions derived from the informed consent documents.

In this study, we build on our previously developed Informed Consent Ontology (ICO), https://github.com/ICO-ontology/ICO. Intended for use as a reference ontology, ICO represents processes and information entities pertaining to informed consent in biomedical investigations. ICO is based on the upper level Basic Formal Ontology (BFO), and adheres to the OBO Foundry framework for authoring and editing BFO-based ontologies[4]; it is licensed under the Creative Commons Attribution 4.0 International Public License(CC BY 4.0). The August 2020 ICO release includes a full alignment with DUO and a new set of 'information content entities' that reflect categories of information needed to fully express information about permissions, i.e. permission-directives as well as relevant-to-permission information. The two classes of 'directive information content entity' and 'designative information content entity' are hierarchical, with subclasses of 'permission directive' and 'permission condition directive', and 'designated actor', 'designated action' and 'designated object'. Each of the subclasses has a small number of additional hierarchically modeled classes. We define a 'permission-directive' as a directive information content entity that prescribes an allowable action, where that action is otherwise impermissible. Permission directives are usually expressed in IC documents as statement(s) that, upon signature of the consent form, authorizes actions by actors that would otherwise not be allowable. Phrases such as "I consent to", or "I agree to" may be indications of permission-directives. More specific information about what action may occur, when it may occur, why it will occur, and any conditions under which it may occur is commonly found in other clauses, and often not co-located with the permission-directive.

**Semantic Web Rule Language**. Semantic Web Rules Language (SWRL)[5] is a W3C semantic web language extension to implement predicate logic on variables. SWRL's benefit in relation to ontologies is to enable expressive machine-based rule reasoning for instance-level data and add additional expressiveness on top of the Web Ontology Language (OWL)[6] . The composition of SWRL contains two conditional propositions - antecedent condition (AC) and consequent condition (CC). The AC is a conditional body proposition that describes the "if" aspect of a rule statement and the CC is the head proposition that describes the "then" condition. Basically, the completeness of a SWRL is contingent on whether the conditions described in the body (AC) and the head (CC) conditions are satisfied. The expression below shows an example of a simple SWRL statement.

*Person(?a) ^ of_age (?a, ?age) ^ swrlb:greaterThan(?age,18) ^  residesInCountry(?a, United States) -> Adult(?a)*

For the simplified example described, a rule statement for determining legal adulthood is complete if only if the entity (?a) is a person that has an age greater than or equal to the value of 18 and resides in the United States. Each component within the statement is an atom which is the core building block of the statement. These SWRL atoms describe the classes (Person), individuals (a, age), data values (18), and other built in utility arguments. For this paper we introduce the possibility of expressing the complexity and execution of permissions in IC documents using SWRL and the ICO.

**Research Objective:** In an effort to extend the ICO, we intend to use SWRL to enhance the expression and application of informed consent . The application of SWRL could further enhance the expressive power of ICO by mapping specific data entities in the informed consent process (actors, data, agreement, etc.) and operationalize it for the software system. ***We presume that ICO through the extension of SWRL can express and operationalize permissions embedded in informed consent documents****.* The outcome of this study would demonstrate the practicality of using SWRL to extend the operational use of ICO for software systems.

**Methods**

To determine if our approach could successfully represent contemporary permission constructs we examined a single IC document, the sample primary IC document from the NIH "All of Us" research program, dated June 20, 2018. The All of Us program seeks to recruit one million or more diverse participants to a large observational trial with the end goal of supporting a variety of use cases within precision medicine (https://allofus.nih.gov/). When signed, an "All of Us" IC document authorizes the collection, retrieval, sharing, and reuse of participant's basic data, health data, physical measurements, biological specimens, fitness trackers and additional data from sources such as health registries, pharmacies, and claims data. Content from electronic health records can be used if the HIPAA authorization form is signed. We selected this IC document for this study because of the scope and familiarity of the "All of Us" study, and because in our earlier review of IC documents retrieved from ClinicalTrials.gov, the readability of the "All of Us" document was at an easy to read level, and thus did not present our study with ambiguous language. We did not include the separate HIPAA authorization form that is required if participants choose to allow access to their electronic health record data.

The All of Us IC document includes 270 sentences. Only one was a permission-directive, 130 (48%) were relevant-to-permission sentences, and 139 (52%) sentences categorized as "other," i.e. important information but not relevant to the purpose of this study. A single permission-directive was identified: "I freely and willingly choose to take part in the All of Us Research Program.". Examples of permission-relevant sentences included the following:

- *"These are results that could be used by a healthcare provider to take better care of you. For example, if any of your physical measurements are outside of what we would expect, we will tell you so you can follow-up with your healthcare provider."*
- *"If I give a blood, urine, or saliva sample, it will be stored at the All of Us biobank. This includes my DNA."*
- *"Information that researchers learn by studying my samples will be stored in the All of Us databases."*
- *"Researchers will do studies using the All of Us databases and biobank. Their research may be on nearly any topic."*

With each permission there are one or more consent directives that link the designated instance data for designated actors, designated actions, designated purposes, and designated objects. Essentially a data instance of a consent directive would link to instance data, and related entities from the consent document. As we will show later, through semantic rules, additional links based on the few links described will be inferred to show implications of agreeing to the consent. *Definition 1* of our model elucidates the notion that IC documents may contain various permissions and that each permission has a consent directive entity. *Definition 2* further elaborates on our model with designated entities of actors, action, purpose, and object.

**Definition 1.** For any informed consent form $icf$ there exist a set of permissions $P$ belonging to an informed consent form $icf$. Also, there exists one consent directive $cd$ for every permission $P$.

$$\{P_1, P_2, P_3, \ldots, P_n\} \in icf$$

$$cd_n \in P_n \ where \ n > 0$$

**Definition 2.** We define designated entities $DE$ as a set of designated action $da$, designated actor $dr$, designated purpose $dp$, and designated object $do$, that has a relation $rel^{about}$ with a consent directive $cd$ that describe $A$.

$$DE = \{da, dr, dp, do\}$$

$$rel^{about}(cd_n, DE_m) = A; \ where \ n > 0 \ and \ m > 0$$

Each permission describes probable *designated actors* that include the subject (the person signing the consent form – the designated permitting actor), the institutions and organization that may be involved – *e.g.,* National Institute of Health, some research organization, All of Us initiative, and the research team members including the principal investigator (the designated permitted actors). Additionally, *designated object* for each of the models are described if the permission refers to subject's data – *e.g.,* general health information, personal health information, or some biospecimen. The models also include the *designated purpose* and *designated action* that expresses motivation (research, therapy, etc.) and the general action on the *designated object* for which permission is being given (disclosure of data, data collection, etc.).

The permission includes *planned processes* that are described in the informed consent ontology where the designated entities link either explicitly or implicitly (through machine inference). For the former and depending on the expression, there is a defined implication that consenting is *act of authorizing* some type of permission to another actor. For the implicit links to the *planned process*, there are allusions to future processes if the subject consents – *e.g.,* granting permission to the researchers to use the shared data, agreeing to inform the patient of change of care, storing the subject's data, etc. These implicit inferences, since the entities are modeled as instance data, will be derived from Semantic Web Rule Languages (SWRL) to be discussed in the next section. *Definition 3* explains that for a permission, there are predicates (defined as *A*) that are members of a permission of an informed consent. These predicates make references to *planned process*(es), which can be inferred.

**Definition 3.** Every permission *P* of an informed consent form entails a subset *A* that produces an allusion → to some planned process *pp*. Essentially, any derivative of *A* is an allusion → to some planned process *pp*.

$$P \ni \{\{A_1, A_2, A_3 \ldots, A_n \to pp\}\}$$

$$\text{where } A \iff rel^{about}(cd, \{da_m, dr_m, dp_m, do_m\}), m > 0$$

$$\therefore rel^{about}(cd, \{da_m, dr_m, dp_m, do_m\}) \to pp$$

With our general defined model of abstracting permissions from informed consent documents, we developed SWRL expressions that can be reasoned on the available entity data extracted from the permissions to infer entities that may be implied when agreeing to participate in a study. We chose the aforementioned four permissions from All of US document to test our approach.

**Results**

For each of the aforementioned examples (Use cases 1 through 4), we produced individual SWRL expressions to generate implied agreement for procedures based on entity data derived from the informed consent document. We populated the ICO ontology with instance level data associated with probable entities in Protégé 5.5 [7]. We executed the SWRL rules and reviewed the generated inferences to assess execution of the rule in an iterative process to refinement. We utilized the built-in Hermit reasoner of Protégé which supports the generation of inferred axioms from SWRL[8].

**Use Case 1**. This use case involves understanding the process of participating in the informed consent process for collection of participant's health data (*designated process of collection* and *designated health information*) for research (*designated biomedical research purpose*). The underlying assumption that once the subject agrees to this permission (*consent directive*) there is presumed permission (*legally effective consenting*) and that the researchers have informed the participant of the study process (*explaining to participant candidate in informed consent process*) and met the baseline requirements for informing, which are inferred and reasoned through the ontology. Following the participant (designated permitting actor) providing consent, the research actors become *designated permitted actors* by inference. Figure 1 models Use Case 1 permission, and Code Listing 1 shows the SWRL expression to execute our use case.

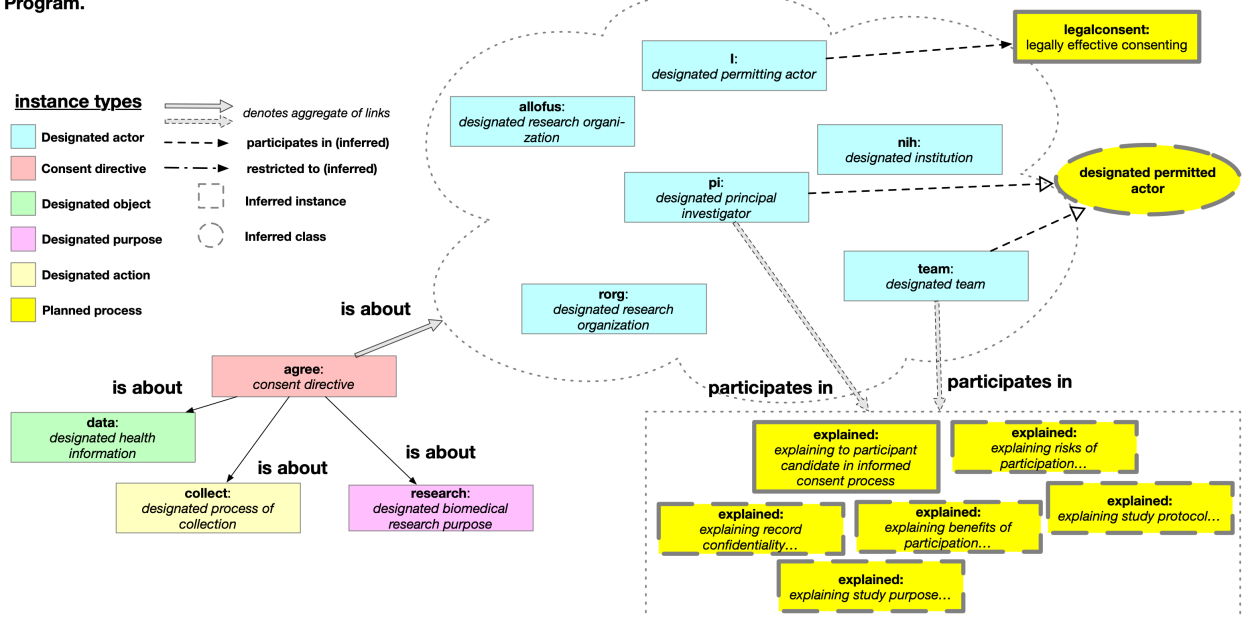

**Figure 1.** Use Case 1's instance-level ontological model

obo:ICO_0000322(?agree) ^ obo:ICO_0000370(?data) ^ obo:IAO_0000136(?agree, ?data) ^obo:ICO_0000332(?collect) ^ obo:IAO_0000136(?agree, ?collect) ^obo:ICO_0000345(?research) ^ obo:IAO_0000136(?agree, ?research) ^obo:ICO_0000395(?rorg) ^ obo:IAO_0000136(?agree, ?rorg) ^obo:ICO_0000398(?I) ^ obo:IAO_0000136(?agree, ?I) ^obo:ICO_0000382(?pi) ^ obo:IAO_0000136(?agree, ?pi) ^obo:ICO_0000381(?nih) ^ obo:IAO_0000136(?agree, ?nih) ^obo:ICO_0000396(?team) ^ obo:IAO_0000136(?agree, ?team) ^obo:ICO_0000142(?legalconsent) ^ obo:ICO_0000154(?explained)
->
obo:RO_0000056(?I, ?legalconsent) ^ obo:RO_0000056(?pi, ?explained)^obo:RO_0000056(?team, ?explained) ^ obo:ICO_0000378(?pi) ^obo:ICO_0000378(?team)^obo:ICO_0000121(?explained)^obo:ICO_0000135(?explained)^obo:ICO_0000128(?explained)^obo:ICO_0000117(?explained)^obo:ICO_0000105(?explained)

**Listing 1.** Use Case 1's SWRL expression.

The SWRL code from Listing 1 successfully executed the embedded rules (See Figure 2). From the figure, it shows:

1. The participant (designated permitting actor), through inference and formally agreeing to the permission, participating in (*participates in*) the legally effective consenting ("legalconsent").
2. The research team consisting of the primary investigator ("pi") and the team ("team") are inferred as *designated permitted actors* as a result of the subject executing consent to the permission. In addition there's an expectation of participation in (*participates in*) explaining details of the study as expressed in the informed consent form (see 4).
3. *participates in* is an object property of ICO that has an inversed version, has participant. As a result, the ontology reasoned *has participant* for the subject that consented to the permission ("I") with the *legally effective consenting* (a *planned process*) as the domain.
4. Similar to 3, the *planned process* of *explaining to potential participant through study informing process* inferred has participant for the research team and the PI ("pi"). There was also a set of *planned processes* relating to explaining the study information in the informed consent form that was reasoned. Through implication, these covered the informed consent requirements when discussing the study.

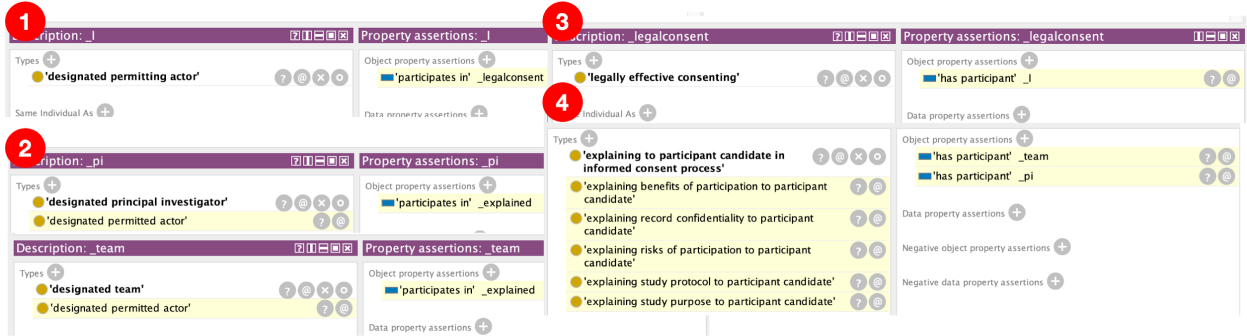

**Figure 2.** Series of screenshots showing inferences generated from SWRL for Use Case 1.

**Use case 2**. The second use case outlines the permitting actor's agreement (after consent) that, if participant data were to have indications that were abnormal, then the researchers' responsibility is to inform the participant about results in order that he/she may seek a health care provider; the expectation is that participant will follow through. This use case model infers (*participates in*) that the *consenting permitting actor*, is likely to seek *standard medical treatment* and infers that the researchers, both the primary investigator and the team will carry out (*participates in*) the informing process (*act of informing*). Similar to first case, the researchers, upon the subject's agreement, will be *designated permitted actors*. Figure 2 visualizes this use case model's execution of the permission's agreement by the subject and Listing 2 shows the corresponding SWRL code.

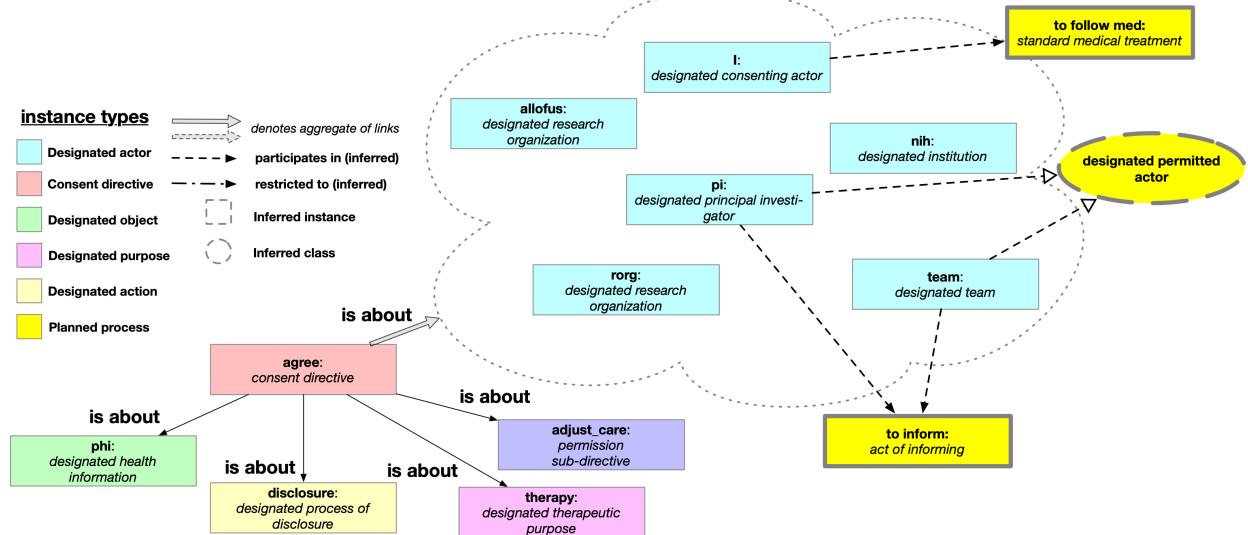

**Figure 3.** Use Case 2's instance-level ontological model.

```
obo:ICO_0000322(?agree) ^ obo:ICO_0000365(?adjust_care) ^ obo:IAO_0000136(?agree, ?adjust_care) ^ obo:ICO_0000370(?phi) ^
obo:IAO_0000136(?agree, ?phi) ^ obo:ICO_0000354(?therapy) ^ obo:IAO_0000136(?agree,?therapy) ^ obo:ICO_0000334(?disclosure) ^
obo:IAO_0000136(?agree, ?disclosure) ^obo:ICO_0000398(?I) ^ obo:IAO_0000136(?agree, ?I) ^ obo:ICO_0000395(?allofus) ^
obo:IAO_0000136(?agree, ?allofus) ^obo:ICO_0000382(?pi) ^obo:IAO_0000136(?agree,?pi)^ obo:ICO_0000395(?rorg) ^ obo:IAO_0000136(?agree,
?rorg) ^ obo:ICO_0000381(?nih) ^ obo:IAO_0000136(?agree, ?nih) ^ obo:ICO_0000396(?team) ^ obo:IAO_0000136(?agree, ?team)
^obo:ICO_0000269(?tocommunicate) ^ obo:ICO_0000110(?medical)
-> obo:RO_0000056(?I,?medical) ^ obo:RO_0000056(?pi,?tocommunicate) ^obo:RO_0000056(?team, ?tocommunicate) ^ obo:ICO_0000378(?pi)
^obo:ICO_0000378(?team)
```

**Listing 2.** Use Case 2's SWRL Expression.

From the execution of the SWRL coding, the following figure (Figure 4) shows the various inferences :

1. The designated permitting actor (the subject who agreed to the permission) is inferred to comply in participating in (*participates in*) following up with medical care if the subject's data is abnormal.
2. The investigation team of the primary investigator ("pi") and cohort ("team") is inferred in participating in informing ("toinform"). This also includes assigning them as *designated permitted actor(s)*.

3. The SWRL execution also includes inferring the inverse of *participates in* (*has participant*) for the *designated permitted actors* and *designated permitting actor*.

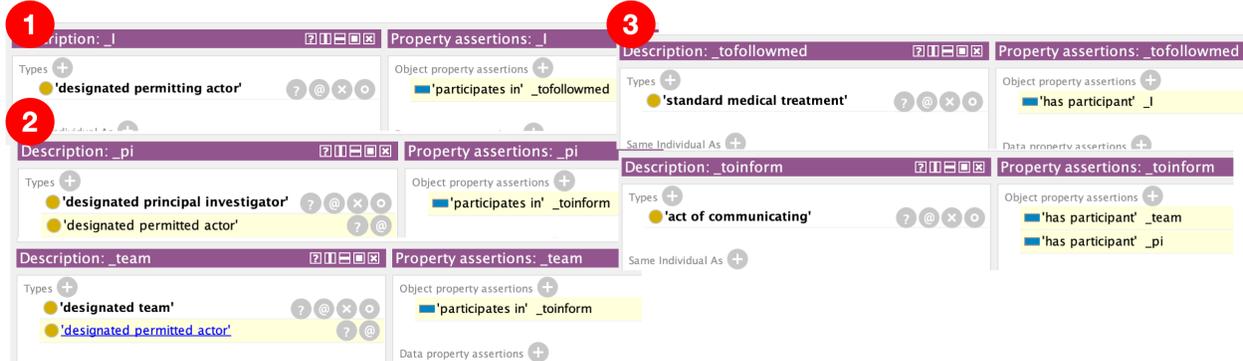

**Figure 4.** Series of screenshots showing inferences generated through SWRL for Use Case 2.

**Use Case 3**. The third use case describes a model where upon the subject's execution of agreement, it permits the execution of authorizing (*act of authorizing*) the sharing of his/her biospecimen data (*act of data sharing*) and the storage of subject's biospecimen data (*act of storing a specimen*) by the inferred *designated permitted actors* of the primary investigator and his/her team. The data to be stored is restricted (via *restricted to*) to just the urine, blood, and saliva. This use case's SWRL code is described in Listing 3.

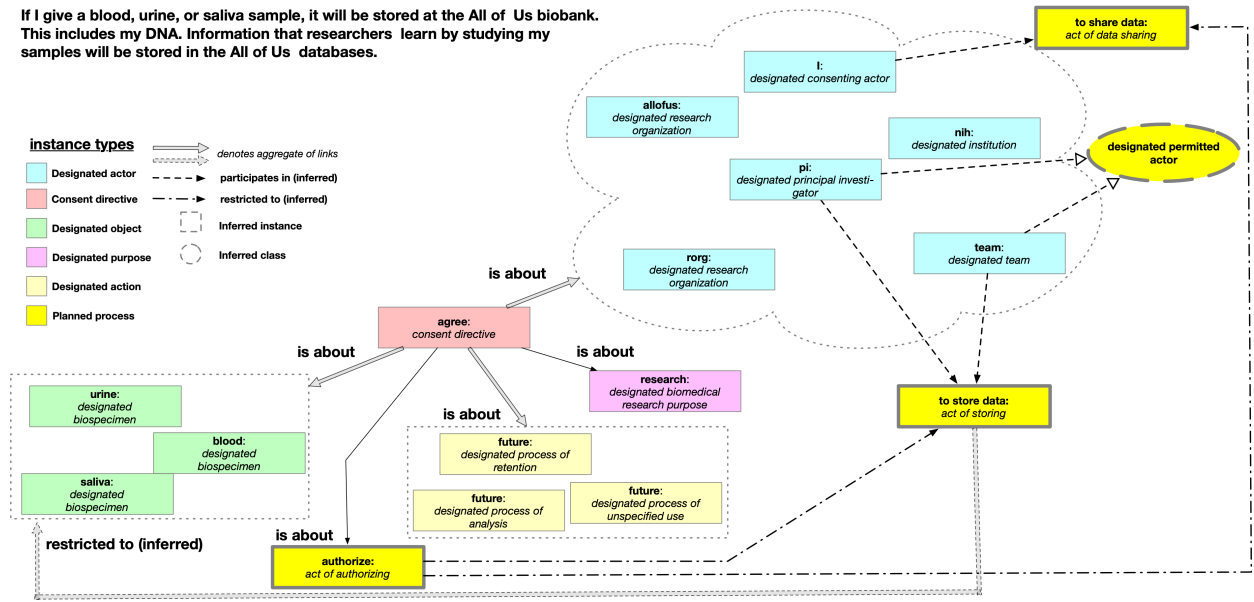

**Figure 5.** Use Case 3's instance-level ontological model.

obo:ICO_0000322(?agree) ^ obo:ICO_0000375(?biospecimen)^obo:IAO_0000136(?agree, ?biospecimen)^obo:ICO_0000339(?future)^obo:ICO_0000330(?future)^obo:ICO_0000336(?future)^obo:IAO_0000136(?agree, ?future) ^obo:ICO_0000345(?research) ^ obo:IAO_0000136(?agree, ?research) ^ obo:ICO_0000046(?authorize) ^ obo:IAO_0000136(?agree, ?authorize) ^ obo:ICO_0000395(?rorg) ^ obo:IAO_0000136(?agree, ?rorg) ^obo:ICO_0000398(?I) ^ obo:IAO_0000136(?agree, ?I) ^ obo:ICO_0000382(?pi) ^ obo:IAO_0000136(?agree, ?pi) ^obo:ICO_0000381(?nih) ^ obo:IAO_0000136(?agree, ?nih) ^obo:ICO_0000396(?team) ^ obo:IAO_0000136(?agree, ?team) ^obo:ICO_0000228(?tosharedata) ^ obo:ICO_0000060(?tostoredata)
->
obo:RO_0000056(?I, ?tosharedata) ^ obo:RO_0000056(?pi, ?tostoredata) ^obo:RO_0000056(?team, ?tostoredata) ^ obo:ICO_0000378(?pi) ^ obo:ICO_0000378(?team) ^ obo:DUO_0000010(?authorize, ?tosharedata) ^ obo:DUO_0000010(?authorize, ?tostoredata) ^ obo:DUO_0000010(?tostoredata, ?biospecimen)

**Listing 3.** Use Case 3's SWRL expression

The results of SWRL expression from Listing 3 are presented in the figure below (Figure 6). As a result of the research subject executing consent here are the inferred expectations:

1. The research subject who consented to sharing data *participates in act of data sharing* ("tosharedata").
2. The subject authorization is restricted to storing and sharing of his data – "tostoredata" and "tosharedata".

3. Research team members involved are inferred to be *designated permitted actor(s)* and are inferred with the expectation to store the subject's data with *participates in*.
4. From the participates in inverse feature, the ontology infers has participant for *act of data sharing* (for the subject "I")…
5. …and for *act of storing a specimen* (for the "pi" and "team" – researchers). Additionally, the storage of data is restricted to just the participant's saliva, blood, and urine with an inference of *is restricted to*.

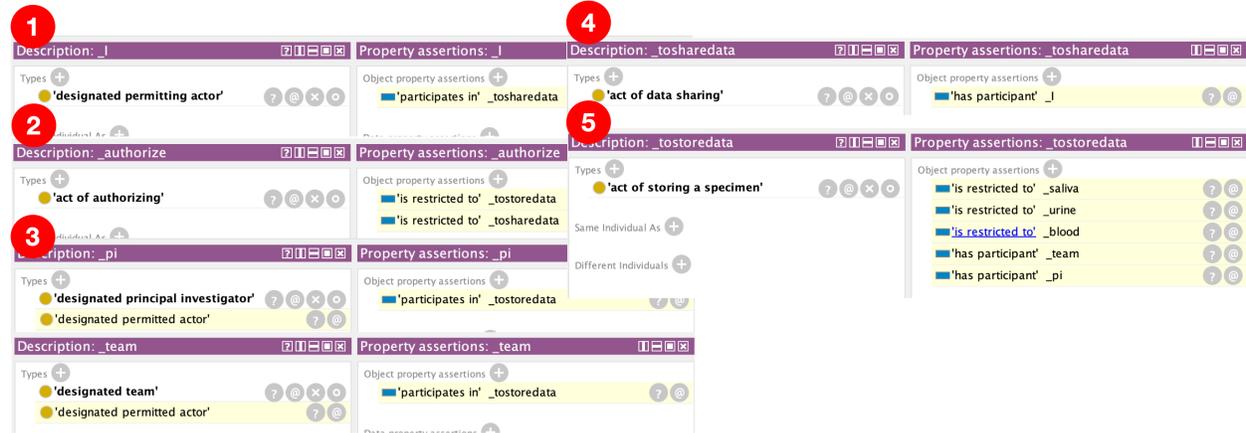

**Figure 6.** Series of screenshot showing inferences generated through SWRL for Use Case 3.

**Use case 4.** The model for this use case expresses conditions that indicate restrictions based on the data available from the participant (urine, blood, saliva of type *designated biospecimen*) and their use will be for any research topic in the future (*designated process of unspecified use* and *designated research purpose*). Structurally, the abstraction for this model shares some similarity with Use Case 3. If the research participant consents, then the permitting actor authorizes (*act of authorizing*) the use of the data by the researchers (inferred as *designated permitted actors*) is restricted to (by inference) the subject's urine, blood, and saliva. The code listing (Code Listing 4) shares some similarity with the previous that was executed.

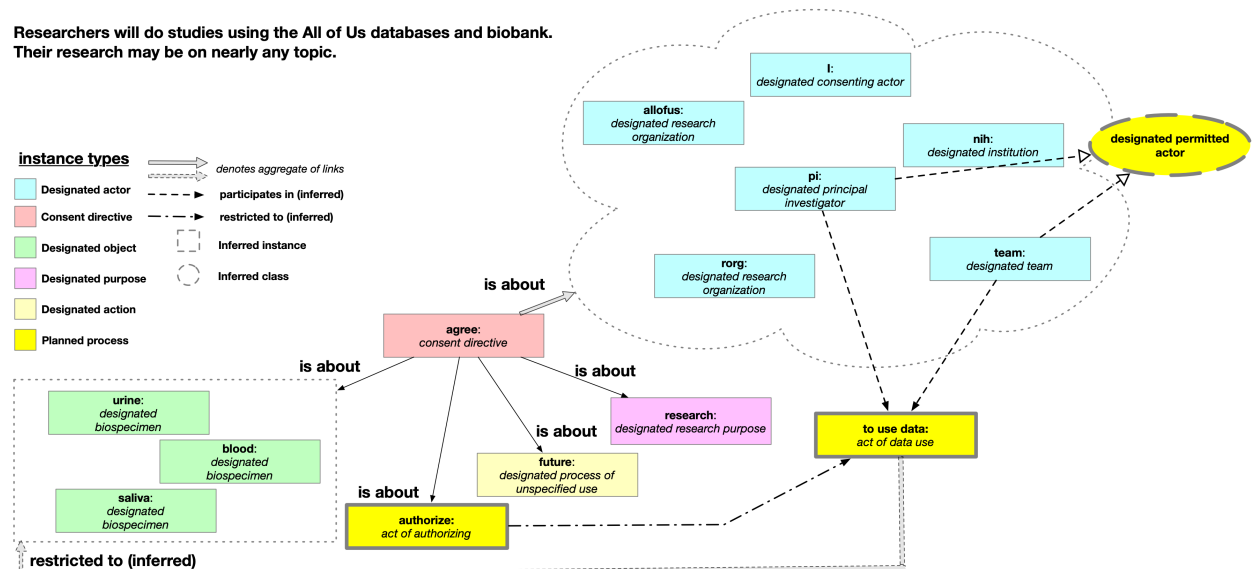

**Figure 7.** Use Case 4's instance level ontological model.

obo:ICO_0000322(?agree) ^ obo:ICO_0000375(?biospecimen)^obo:IAO_0000136(?agree, ?biospecimen)^obo:ICO_0000336(?future)^obo:IAO_0000136(?agree, ?future) ^obo:ICO_0000344(?research)^obo:IAO_0000136(?agree, ?research) ^ obo:ICO_0000046(?authorize) ^ obo:IAO_0000136(?agree, ?authorize) ^ obo:ICO_0000395(?rorg) ^ obo:IAO_0000136(?agree, ?rorg) ^obo:ICO_0000398(?I) ^ obo:IAO_0000136(?agree, ?I) ^obo:ICO_0000382(?pi) ^ obo:IAO_0000136(?agree, ?pi) ^obo:ICO_0000381(?nih) ^ obo:IAO_0000136(?agree, ?nih) ^obo:ICO_0000396(?team) ^ obo:IAO_0000136(?agree, ?team) ^obo:ICO_0000421(?tousedata)
->
obo:RO_0000056(?pi, ?tousedata) ^ obo:RO_0000056(?team, ?tousedata) ^ obo:ICO_0000378(?pi) ^ obo:ICO_0000378(?team) ^ obo:DUO_0000010(?authorize, ?tousedata) ^ obo:DUO_0000010(?tousedata, ?biospecimen)

**Listing 4.** Use Case 4's SWRL expression

The following figure (Figure 8) shows the successful execution of the use case's model from Listing 4. In the figure we show:

1. The act of authorizing by the subject *is restricted to the act of using study participant data.*
2. Similar to the other example use cases, the research team is inferred to be designated permitted actor(s) and they will *participate in* using the data (*act of data using study participant data* ).
3. Like Use Case 3, the act of using study participant data is restricted to subject's blood, urine, and saliva, and through the inverse of *participates in,* expresses the inference that this act involves the research team of the primary investigator ("pi") and his/her research team ("team")

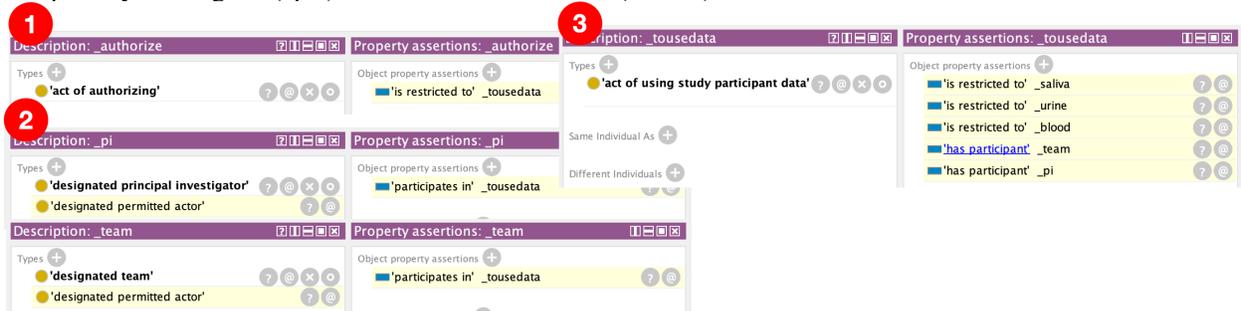

**Figure 8**. Series of screenshots showing inferences generated through SWRL for Use Case 4.

**Discussion**

In this project, we demonstrate the utilization of the Semantic Web Rule Language (SWRL), an extension of OWL, to provide inferencing of entity instance data from informed consent documents. The use of SWRL could effectively enhance the application of ICO to manage complex and essential IC data that must be linked for understanding permissions. Moreover, this work could be potentially integrated into software systems. In this way, we address the gap in establishing traceability from details included in the consent form to repositories and other systems.

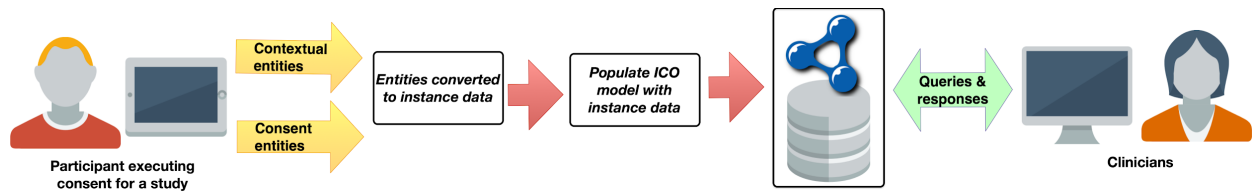

**Figure 9.** Software system for managing and querying informed consent showing the integration.

While much of what we presented was implemented through the Protégé environment, with the SWRLAPI [9] combined with the ICO model, one can potentially create custom software integration to be incorporated into software systems, as shown in Figure 9. The study participant, through an interface, confirms his/her agreement to participate in a study by signing an IC document. The data from the IC document, along with the contextual and meta-data information that are not explicitly stated, are transferred from the software interface to the system. The system software converts these data to instance data. This instance data is added and labeled to ICO, and with the assistance of SWRL extensions, reasoning is performed on the data. The consent data (including the reasoned consent data) can later be queried by clinicians.

From our four use cases we demonstrated the utility of our proposed method. Each use case demonstrated not only translating the permission to a computable format using ontology language (OWL and SWRL), but they also demonstrated predictability inferencing, using ontology-based reasoning to generate links to entities that may not be explicitly evoked in the consent form (e.g., identifying designated actors).

Several limitations are noted. The rules and the ontological translations are only examples, and used to show proof of concept; they are not comprehensive or complete since we limited our sample to one IC document with one permission-directive. In addition, we have only used the "is_about" relation; our goal here was to identify information content that could be tagged and serve as a foundation for use in computable semantics such as inferencing operations.

There were a handful of concepts that could not be precisely expressed, due to a lack of coverage of minor entities. Some of this apparent lack of coverage in ICO may result from the need to import content from other BFO-based ontologies in the OBO Foundry. These cases need to be enumerated and then used to either identify gaps in ICO itself, or negotiate inclusion in other ontologies in the OBO Foundry. ICO provides a general model of informed consent and does not include in its scope the modeling of regulatory concepts; there is a need to support the creation of SWRL rules to help extend legal, regulatory, and policy expressions. One probable future goal is to automate the translation of SWRL rules from informed consent documents. However, a more achievable and transparent method would be a visual authoring tool to create the SWRL expressions that align with the permissions in the informed consent document. We are encouraged from our experience in authoring the expressions based on the ontology and the SWRL rules, that this could be generalized for most of the informed consent documents as there are consistent semantic structures involved in accurately expressing permissions. Our group is pursuing the creation and validation of such models. Finally, the focus of the work we presented is permissions embedded in the informed consent document. Our observation in the use cases are of similar and repeated abstractions, and contexts that were on the document or meta-level but not explicitly stated on the permission level. Part of our future direction is to approach the ontological expression on a "document level" that would collapse or merge the abstraction to avoid repetition and redundancy.

**Conclusion**

Through the use of semantic web rule language (SWRL) that enables machine-level reasoning on the instance-level data of ICO, we were able to demonstrate our proof of concept method to link and generate information from entities derived from four permissions in the All of Us informed consent document. Our efforts allowed us to extend the application use of ICO to be potentially integrated with software systems to manage complex authorization information that can later be supported by queries. Our future direction is to work towards formalizing a set of generalized rule extensions for ICO to accurately express document-level permission entities from informed consent.

**Acknowledgements.** This research was supported by the NIH/NCI under Contract Number 75N91020C00017 (Manion PI), and builds off work supported by an award from the Michigan Institute for Data Science (Harris PI), and NIH/NHGRI U01 HG009454 (Tao PI); and by the National Institute of Allergy and Infectious Diseases of the National Institutes of Health under Award No, R01AI130460 (Tao, PI) and R01AI130460-03S1 (Tao, PI).